\documentclass[ALICE,manyauthors]{cernphprep}

\usepackage{hyperref}
\usepackage{lineno}

\usepackage{graphicx,epsfig}
\usepackage[normalem]{ulem}
\usepackage{color}
\usepackage{indentfirst}
\usepackage{xspace}
\usepackage{wasysym}
\usepackage{pdflscape}
\usepackage{multirow}

\usepackage[comma,square,numbers,sort&compress]{natbib}

\graphicspath{{Figures/}}

\modulolinenumbers[5]

\newcommand{\tocsecindent}{\hspace{6mm}}
\newcommand{\quotes}[1]{``#1''}

\newcommand{\pA}{p--A\xspace}
\newcommand{\pPb}{p--Pb\xspace}
\newcommand{\Pbp}{Pb--p\xspace}
\newcommand{\PbPb}{\mbox{Pb--Pb}\xspace}
\newcommand{\AAcol}{\mbox{A--A}\xspace}
\newcommand{\pp}{\mbox{pp}\xspace}
\newcommand{\dAu}{d-Au\xspace}
\newcommand{\jpsi}{\ensuremath{\mathrm{J}/\psi}\xspace}
\newcommand{\jpsitoll}{\ensuremath{\mathrm{J}/\psi\rightarrow l^+l^-}\xspace}

\newcommand{\psip}{\ensuremath{\psi\mathrm{(2S)}}\xspace}
\newcommand{\chic}{\ensuremath{\chi_{c}}\xspace}
\newcommand{\pt}{\ensuremath{p_{\mathrm{T}}}\xspace}

\newcommand{\xbj}{\ensuremath{x_{\mathrm{Bj}}}\xspace}
\newcommand{\gevc}{\ensuremath{\mathrm{GeV}/c}\xspace}

\newcommand{\gevcsqr}{\ensuremath{\mathrm{GeV}/c^{2}}\xspace}

\newcommand{\ylab}{\ensuremath{y_\mathrm{lab}}\xspace}
\newcommand{\ycms}{\ensuremath{y_\mathrm{cms}}\xspace}
\newcommand{\etag}{\ensuremath{\eta}\xspace}

\newcommand{\dedx}{\ensuremath{{\mathrm d}E/{\mathrm d}x}\xspace}
\newcommand{\Rpa}{\ensuremath{R_{\mathrm{pPb}}}\xspace}

\newcommand{\Qpa}{\ensuremath{Q_{\mathrm{pPb}}}\xspace}
\newcommand{\Qpainc}{\ensuremath{Q_{\mathrm{pPb}}^{\mathrm{incl}}}\xspace}
\newcommand{\Qpab}{\ensuremath{Q_{\mathrm{pPb}}^{\mathrm{non-prompt}}}\xspace}
\newcommand{\Qpap}{\ensuremath{Q_{\mathrm{pPb}}^{\mathrm{prompt}}}\xspace}

\newcommand{\avTpa}        {\ensuremath{\langle T_\mathrm{pPb} \rangle}\xspace}
\newcommand{\avTpaMB}      {\ensuremath{\langle T_\mathrm{pPb}^\mathrm{MB} \rangle}\xspace}
\newcommand{\avNpart}      {\ensuremath{\langle N_\mathrm{part} \rangle}\xspace}
\newcommand{\avNpartMB}    {\ensuremath{\langle N_\mathrm{part}^\mathrm{MB} \rangle}\xspace}

\newcommand{\avNcoll}      {\ensuremath{\langle N_\mathrm{coll} \rangle}\xspace}
\newcommand{\avNcollMB}    {\ensuremath{\langle N_\mathrm{coll}^\mathrm{MB} \rangle}\xspace}
\newcommand{\avNcollmult}      {\ensuremath{\langle N_\mathrm{coll}^\mathrm{mult} \rangle}\xspace}
\newcommand{\avncollmult}      {\ensuremath{\langle N_\mathrm{coll}^\mathrm{mult} \rangle}\xspace}
\newcommand{\avTpamult}      {\ensuremath{\langle T_\mathrm{pPb}^\mathrm{mult} \rangle}\xspace}

\newcommand{\avTpPbmult}      {\ensuremath{\langle T_\mathrm{pPb}^\mathrm{mult} \rangle}\xspace}
\newcommand{\sigpp}{\ensuremath{\sigma_{\mathrm{pp}}}\xspace}
\newcommand{\fnorm}{\ensuremath{F_{2\mu/\mathrm{MB}}}\xspace}

\newcommand{\nmb}{\ensuremath{N_{\mathrm{MB}}}\xspace}
\newcommand{\sigmb}{\ensuremath{\sigma_{\mathrm{MB}}}\xspace}
\newcommand{\meanpt}{\ensuremath{\langle p_{ {\mathrm T} } \rangle}\xspace}
\newcommand{\meanptsqr}{\ensuremath{\langle p^2_{ {\mathrm T} } \rangle}\xspace}
\newcommand{\deltameanptsqr}{\ensuremath{\Delta \langle p^2_{ {\mathrm T} } \rangle}\xspace}
\newcommand{\s}{\ensuremath{\sqrt{s}}\xspace}
\newcommand{\snn}{\ensuremath{\sqrt{s_{\mathrm{NN}}}}\xspace}
\newcommand{\mumu}{\ensuremath{\mu^+\mu^-}\xspace}
\newcommand{\elel}{\ensuremath{e^+e^-}\xspace}
\newcommand{\ccbar}{\ensuremath{c\overline{c}}\xspace}
\newcommand{\ncollmult}{\ensuremath{N_{\rm coll}^{\rm mult}}\xspace}
\newcommand{\centrality}{centrality\xspace}
\newcommand{\cerenkov}{Cherenkov\xspace}
\newcommand{\fb}{\ensuremath{f_{B}}\xspace}
\newcommand{\mmumu}{\ensuremath{m_{\mu^+\mu^-}}\xspace}

\renewcommand{\refname}{References}

\begin{document}

\begin{titlepage}
\PHyear{2015}
\PHnumber{158}      
\PHdate{June 26}  
%

\title{Centrality dependence of inclusive J/$\mathbf{\psi}$ production \\
in \mbox{p--Pb} collisions at $\mathbf{\sqrt{{\textit s}_{\rm NN}}}$~=~5.02~TeV}
\ShortTitle{Centrality dependence of inclusive J/$\mathbf{\psi}$ production in \mbox{p--Pb} at $\mathbf{\sqrt{s_{\rm NN}}}$=5.02 TeV}   

\Collaboration{ALICE Collaboration\thanks{See Appendix~\ref{app:collab} for the list of collaboration members}}
\ShortAuthor{ALICE Collaboration} 

\begin{abstract}
We present a measurement of inclusive \jpsi production in \pPb collisions at $\snn=5.02$~TeV as a function of 
the \centrality of the collision, 
as estimated from the energy deposited in the Zero Degree Calorimeters. 
The measurement is performed with the ALICE detector down to zero transverse momentum, \pt, 
in the backward ($-4.46 < y_{\rm cms} < -2.96$) and forward ($2.03 < y_{\rm cms} < 3.53$) rapidity intervals in the dimuon decay channel and 
in the mid-rapidity region ($-1.37 < y_{\rm cms} < 0.43$) in the dielectron decay channel. The backward and forward rapidity intervals 
correspond to the Pb-going and p-going direction, respectively. 
The \pt-differential \jpsi production cross section at backward and forward rapidity is measured for 
several \centrality classes, together with the corresponding average \pt and $\pt^2$ values. 
The nuclear modification factor is presented as a function of 
\centrality for the three rapidity intervals, and  
as a function of \pt for several \centrality classes at backward and forward rapidity. 
At mid- and forward rapidity, 
the \jpsi yield is suppressed up to $40\%$ compared to 
that in \pp interactions scaled by the number of binary collisions. 
The degree of suppression increases towards central \pPb collisions at forward rapidity,
and with decreasing \pt of the \jpsi. 
At backward rapidity, the nuclear modification factor is compatible with unity within the total uncertainties, 
with an increasing trend from peripheral to central \pPb collisions.
\end{abstract}
\end{titlepage}
\setcounter{page}{2}

\section{Introduction}
\label{sec:intro}

Charmonia, bound states of charm and anti-charm quark pairs, are extensively used 
to study the interplay between the perturbative and the non-perturbative regimes of Quantum 
ChromoDynamics (QCD)~\cite{Brambilla:2010cs}. 
Charmonium production mechanism can be understood as a hard scattering, 
describing the charm anti-charm quark pair production, 
followed 
by the evolution of the pair into a bound state via a non-perturbative process. 
Models such as colour evaporation (CEM)~\cite{Fritzsch:1977ay,Amundson:1996qr}, 
colour singlet (CSM)~\cite{Baier:1981uk} and non-relativistic 
QCD (NRQCD)~\cite{Bodwin:1994jh} are used to describe the charmonium production in hadronic collisions. 
None of these models has so far provided a consistent description of the production 
cross section and polarisation measured in proton-proton (\pp) collisions~\cite{Brambilla:2010cs,Andronic:2015wma}. 
The 1S vector state, the \jpsi meson, is abundantly produced in hadronic collisions at high energy and 
measurable through its leptonic decays.
Its inclusive production contains contributions from direct \jpsi, from decays of higher-mass excited states, 
\psip and \chic, as well as from non-prompt \jpsi, from weak decays of beauty hadrons.

In proton-nucleus (\pA) collisions, several effects related to the nuclear medium and commonly denoted as cold nuclear matter effects (CNM) 
can affect the production of charmonia. 
The Parton Distribution Functions (PDFs) of nucleons bound in nuclei are modified compared 
to those of free nucleons~\cite{Eskola:2009uj,deFlorian:2011fp,Hirai:2007sx}. 
These functions depend, in particular, on the fraction of the nucleon momentum, Bjorken-$x$ (\xbj), 
carried by the probed parton. 
In the collision energy regime typical of the Large Hadron Collider (LHC), charm quark pairs are 
produced mainly via the gluon fusion process. 
The gluon nuclear PDFs (nPDFs) are suppressed at low \xbj ($\xbj\lesssim 0.01$), enhanced at intermediate 
\xbj ($0.01 \lesssim \xbj \lesssim 0.3$) and suppressed again at large \xbj 
($0.35\lesssim \xbj \lesssim 0.7$) compared to those of free nucleons. 
These three kinematic regions are often referred to as the shadowing, anti-shadowing, and EMC regions, respectively. 
Alternatively, at low \xbj, 
the initial colliding nucleus can be described by the 
Colour Glass Condensate (CGC) effective theory~\cite{Kharzeev:2005zr,Fujii:2006ab} 
as a coherent and dense (saturated) gluonic system. 
The kinematic distribution of the produced charm quark pairs may be additionally modified by multiple scattering of the incoming
gluons and/or the quark pairs with the surrounding nuclear medium~\cite{Qiu:1998rz,Fujii:2006ab} or by the 
energy loss via gluon radiation~\cite{Gavin:1991qk,Brodsky:1992nq}. It was also argued that the interference 
between the gluons radiated before and after the hard scattering can lead to important coherent energy loss effects 
at large rapidity in the p-going direction~\cite{Arleo:2012rs}. 
Finally, after their formation, the pre-resonant charm quark pairs or the fully formed resonances may interact with the nucleons 
when passing through the nucleus (nuclear absorption~\cite{Vogt:2010aa}) or with the 
other particles produced in the \pPb collision (comovers~\cite{Ferreiro:2014bia}). 
Consequently, they may lose energy or fragment into open charm meson pairs. 
Due to the short time spent by the charm quark pairs in the nucleus relative to the \jpsi 
formation time at LHC energies, 
the effect of nuclear absorption is expected to be small~\cite{Lourenco:2008,Ferreiro:2013pua}.

Charmonium production was predicted to be suppressed 
in a hot medium with a high density of colour charges, the Quark-Gluon Plasma (QGP), 
as a consequence of the colour screening mechanism~\cite{Matsui:1986dk}. Such a state can be formed in 
ultra-relativistic nucleus-nucleus collisions. At the LHC, where the charm-quark density is large, 
charmonium may also be created via the 
(re)combination of charm quarks either during the deconfined phase~\cite{Thews:2000rj}
or at the phase boundary~\cite{BraunMunzinger:2000px}, when the system has cooled down and hadronisation takes place. 
A suppression of \jpsi production in central nucleus-nucleus (\AAcol) collisions with respect to the one measured in 
\pp collisions scaled by the number of binary nucleon-nucleon collisions was observed at the SPS at 
$\snn \sim20$~GeV~\cite{Alessandro:2004ap,Arnaldi:2007zz,Arnaldi:2009ph}, at 
 RHIC at $\snn =$~39, 62.4 and 200~GeV~\cite{Adare:2011yf,Adamczyk:2012ey,Adare:2012wf,Adamczyk:2013tvk} and at the LHC at 
$\snn = 2.76$~TeV~\cite{ALICEAA:2012,Abelev:2013ila,Adam:2015rba}. 
However, the \jpsi production measurements at the LHC show 
a much smaller suppression of the yields integrated over transverse momentum, \pt, as compared to the results at lower collision energies.
 The differential results also
indicate a smaller degree of suppression at $\pt < 3$ \gevc than at higher \pt~\cite{ALICEAA:2013,Chatrchyan:2012np},
at mid- and forward rapidity, in agreement with the expectations from (re)combination models~\cite{Zhao:2011cv,Zhou:2013aea}.
Although, qualitatively, the \PbPb measurements by themselves give a strong indication that the (re)combination effect
plays a significant role in the \jpsi production at LHC energies, the quantitative understanding of 
the involved mechanisms requires a good knowledge of the underlying CNM effects.

The \jpsi production in proton- or deuteron-nucleus collisions was studied at fixed-target  
(SPS~\cite{Alessandro:2006jt,Arnaldi:2010ky}, HERA~\cite{Abt:2008ya}, Tevatron~\cite{Leitch:1999ea}) 
and collider experiments (RHIC~\cite{Adare:2012qf}, LHC~\cite{ALICEpA:2013,Aaij:2013zxa,Adam:2015iga,Aad:2015ddl,ATLASAA:2010}). 
At the LHC, a suppression of the \jpsi production in \pPb collisions with respect to binary-scaled \pp production 
has been observed for $\pt < 5$ GeV/c at large rapidity in the p-going direction and at mid-rapidity, 
 while the measurements at high \pt as well as at large rapidity in the Pb-going direction are consistent with no suppression. 
The results are in fair agreement with models based on shadowing or coherent energy loss. While the \jpsi suppression 
at forward rapidity is overestimated by an early CGC calculation~\cite{Fujii:2013gxa}, recent calculations~\cite{Ma:2015sia,Ducloue:2015gfa} 
are in better agreement with the data. 
The various CNM effects described above should be enhanced at small impact parameters of the collision and thus towards the most central \pPb collisions. Hence, differential measurements as a function of the \pPb collision centrality are essential to further constrain the models, in particular their dependence on the impact parameter of the collision.

In this paper, we report on new results in \pPb collisions at \snn = 5.02 TeV for inclusive \jpsi production, 
measured at backward ($-4.46 < \ycms < -2.96$) and forward ($2.03 < \ycms < 3.53 $) center-of-mass 
rapidity, \ycms, in the \mumu decay channel, and at mid-rapidity ($-1.37 < \ycms < 0.43$) in the \elel decay channel. 
Previous measurements have been carried out as a function of rapidity and \pt~\cite{ALICEpA:2013,Aaij:2013zxa,Adam:2015iga,Aad:2015ddl,ATLASAA:2010}. 
Here, the measurements are performed as a function of the collision \centrality, 
estimated on the basis of the energy deposited in the Zero Degree Calorimeters (ZDC)~\cite{alice-cent}. 
At backward and forward rapidity, the \jpsi cross section is studied as a function of \pt for several \centrality classes. 
The corresponding average values \meanpt, and \meanptsqr, are extracted from the \pt-differential cross sections 
and the \pt broadening, defined as \deltameanptsqr = $\meanptsqr_{\rm pPb} - \meanptsqr_{\rm pp}$, is also discussed. 
The nuclear modification factors are then obtained as a function of \centrality for the three rapidity ranges and at backward and forward rapidity, as a function of \pt for several classes of \centrality.

\section{Detectors and data sets}
\label{sec:genandcent}

The ALICE apparatus and its performance are described in detail in Ref.~\cite{Aamodt:2008zz} and Ref.~\cite{Abelev:2014ffa}, respectively.

Away from the mid-rapidity region, the \jpsi candidates are reconstructed in the \mumu decay channel 
using the muon spectrometer~\cite{Aamodt:2008zz}, covering 
the pseudorapidity range $-4<\eta_{\rm lab}<-2.5$ in the laboratory frame. 
The muon spectrometer includes a dipole magnet with an integrated field of 3 T$\cdot$m, 
five tracking stations comprising two planes of Cathode Pad Chambers each, 
and two trigger stations consisting of two planes of Resistive Plate Chambers each. 
A system of absorbers is used for filtering out the hadrons. The front 
absorber, made of concrete, carbon and steel with a thickness of 4.1~m (10 nuclear interaction lengths, $\lambda_\mathrm{int}$) is installed between 
the interaction region and the muon tracking stations. A second absorber, a 1.2~m thick iron wall (7.2 $\lambda_\mathrm{int}$), 
is located upstream of the trigger stations and 
absorbs secondary hadrons escaping from the front absorber 
and low-momentum muons produced predominantly from $\pi$ and $K$ decays. 
Finally, a conical absorber placed around the beam pipe protects the spectrometer 
from secondary particles produced in interactions of large-\etag primary particles with the beam pipe. 

At mid-rapidity, the \jpsi candidates are measured in the \elel decay channel with the central barrel 
detectors in the pseudorapidity range $|\eta_{\rm lab}|<0.9$. The main subsystems used, the Time Projection Chamber 
(TPC)~\cite{Alme2010316} and the Inner Tracking System (ITS)~\cite{Aamodt:2010aa},
are placed in a solenoidal magnetic field with a strength of 0.5 T. 
The TPC, the main tracking and particle identification device, is a gaseous drift detector with a cylindrical geometry extending
from 85 to 247~cm in the radial direction and 500~cm longitudinally. The particle identification is performed via 
the measurement of the 
specific energy loss, \dedx, in the gas volume. 

The ITS, covering a pseudorapidity range $|\eta_{\rm lab}|<0.9$, consists of 6 layers of silicon detectors placed
at radii ranging from 3.9 to 43~cm relative to the beam axis. The two innermost layers are equipped with 
Silicon Pixel Detectors (SPD). The track segments (tracklets) reconstructed from the hits in the two SPD 
layers are used to reconstruct the interaction vertex position and to reject pile-up events (events with two or 
more simultaneous interactions per bunch crossing). 
The position of the interaction vertex is also determined, with better resolution, 
from the tracks reconstructed in the TPC and the ITS~\cite{Aamodt:2008zz}.  

Two scintillator arrays, V0~\cite{Abbas:2013taa}, placed on both sides of the interaction point (IP) at 
$-3.7<\eta_{\rm lab}<-1.7$\ and $2.8<\eta_{\rm lab}<5.1$,
are used as trigger detectors and to remove beam-induced background. 
They are also used for the measurement of luminosity, along with the T0 detector~\cite{Aamodt:2008zz}, 
consisting of two quartz \cerenkov counters, placed on each side of the IP covering 
the ranges $-3.3<\eta_{\rm lab}<-3.0$ and $4.6<\eta_{\rm lab}<4.9$. 
The Zero Degree Calorimeters (ZDC)~\cite{ALICE:2012aa}, 
located along the beam axis at 112.5~m from the IP on both sides, detect protons and neutrons emitted from the nucleus and are used to estimate 
the centrality of the collision. The neutron calorimeter (ZN) is positioned between the two beam pipes downstream of the first machine 
dipole that separates the beams. The proton calorimeter (ZP) is installed externally to the outgoing 
beam pipe. The ZDCs are also used to remove parasitic \pPb interactions 
displaced from the nominal position. 

The data samples used for the measurements reported in this paper were collected in 2013
in two configurations, obtained by inverting the direction of the p and Pb beams. 
Due to the asymmetry of the energy per nucleon of the p and Pb beams ($E_{\rm p}=4$~TeV and $E_{\rm Pb}/208=1.58$~TeV), 
the nucleon-nucleon center-of-mass system is shifted with respect to the laboratory system by $\Delta y=0.465$ 
in the p-going direction.  
The two beam configurations allow one to measure the \jpsi production in the backward ($-4.46<\ycms<-2.96$) and forward ($2.03<\ycms<3.53$)
 centre-of-mass rapidity (\ycms) regions, corresponding to the Pb-going and the p-going directions, respectively. 
They will be further referred to as 
\Pbp and \pPb, for the first and second case. The dielectron analysis was carried out on a data 
sample corresponding to the \pPb beam configuration, in the mid-rapidity range $-1.37<y_{\rm cms}<0.43$.

The dielectron analysis was performed on a sample of events satisfying a Minimum Bias (MB) trigger condition and 
the dimuon analysis used dimuon-triggered events. 
The MB trigger is defined by a coincidence of the signals from both sides of the V0 detector. 
The efficiency of the MB trigger in selecting non-single diffractive \pPb collisions was estimated to be higher 
than 99\%~\cite{ALICE:2012xs}, with negligible contamination from diffractive collisions. 
The dimuon trigger requires, in addition to the MB condition, the detection of two unlike-sign muon 
candidate tracks in the trigger system of the muon spectrometer. 
This trigger selects muons with a transverse momentum $\pt\gtrsim0.5$~\gevc. This threshold is not sharp in \pt and 
the single-muon trigger efficiency reaches a plateau value of $\sim96\%$ at $\pt\sim1.5$~\gevc. 

In the backward and forward rapidity regions, the measurements are based on a sample of 
$2.1\times10^7$ and $9.3\times10^6$ dimuon-triggered events, respectively. 
The MB interaction rate reached a maximum of 200 kHz, corresponding  
to a maximum pile-up probability of about $3\%$. 
The mid-rapidity data sample consists of $1.1 \times 10^8$ MB-triggered events, collected at a low interaction rate ($\sim10$ kHz) and 
with a fraction of pile-up events lower than $0.6\%$. 
Pile-up of collisions from  different bunch crossings is negligible considering that the 200 ns bunch-crossing spacing 
is larger than the integration time of the ZDC and muon trigger detectors, which are used for the track selection for the dimuon analysis. 
Two independent determinations of the MB trigger cross sections $\sigma_{\rm MB}$ were carried out in the \Pbp and \pPb configurations using van der Meer scans~\cite{vanderMeer:296752}. The corresponding cross sections amount to $\sigma_{\rm MB}^{\rm Pbp}=2.12\pm 0.07~{\rm b}$ and $\sigma_{\rm MB}^{\rm pPb}=2.09\pm 0.07~{\rm b}$, respectively~\cite{Abelev:2014epa}. 
The integrated luminosity is determined as ${\cal L}=\nmb/\sigma_\mathrm{MB}$ where \nmb is the number of 
MB events. The number of MB events corresponding to the dimuon-triggered sample is evaluated as
$\nmb = \fnorm \cdot N_{\rm DIMU}$, where $N_{\rm DIMU}$ is the number of dimuon-triggered events and \fnorm is 
the inverse of the probability of having dimuon-triggered events in a MB data sample. 
The determination of \fnorm is discussed in Sec.~\ref{sec:analysismumu}. 
The integrated luminosity was also independently measured employing the T0 detector. 
The two luminosity measurements agree within better than 1\% throughout the whole \Pbp and \pPb data-taking 
periods~\cite{Abelev:2014epa}. The maximum difference 
is included as an additional uncertainty for $\sigma_{\rm MB}$ and thus in the determination of the 
luminosity uncertainty. 
The integrated luminosity values used for the results in the backward, forward and mid-rapidity regions are $5.81\pm 0.20~{\rm nb}^{-1}$, 
$5.01\pm 0.19~{\rm nb}^{-1}$ and $51.4 \pm 1.9~{\rm \mu b}^{-1}$, respectively. 

The centrality determination in \PbPb collisions is usually based on charged-particle multiplicity, estimated using the V0
signal amplitudes~\cite{Abelev:2013qoq}. However, in \pPb collisions, 
the magnitude of the multiplicity fluctuations at a given impact parameter is comparable 
to the whole dynamic range of the MB multiplicity distribution. 
The fluctuations can be related to the various event topologies 
(e.g. hard collisions with large momentum transfers and/or multiple hard parton-parton interactions, 
which tend to be associated to high multiplicity events, compared to soft collisions without any 
high-\pt particle), detector acceptance effects (jets fragmenting in or out the experimental coverage), 
or other effects, as explained in detail in Ref.~\cite{alice-cent}.
Therefore, a centrality selection based on charged-particle multiplicity may select a
sample of \pPb collisions that contains biases unrelated to the collision geometry. 
In contrast, a centrality selection based on the energy measured with the ZDC in the Pb-going direction, 
deposited by nucleons produced in the nuclear de-excitation processes 
following the collision, or knocked out by wounded nucleons, should not induce such biases. 
The average number of binary nucleon collisions (\avNcoll) or the average nuclear overlap function (\avTpa) 
for a given centrality class, defined by a selected range of energy 
deposited in the Pb-remnant side of ZN, is obtained using the hybrid method described in Ref.~\cite{alice-cent}. 
In this method, the \avNcoll determination relies on the assumption that the charged-particle multiplicity
measured at mid-rapidity is proportional to the number of participant nucleons (\avNpart).
The values of \avNpart for a given ZN-energy class, also noted as ZN class in the following,  
were calculated by scaling the MB value of the number of 
participant nucleons, \avNpartMB, by the ratio of the average charged-particle multiplicities measured 
at mid-rapidity for the considered ZN-energy event classes to the corresponding value in MB collisions. The average number of collisions 
and the average nuclear overlap function were then calculated from \avNpart according to the 
Glauber model~\cite{Miller:2007ri}, which is generally used to calculate geometrical quantities of nuclear collisions. 
From here on these values are denoted as \avncollmult and \avTpPbmult to indicate the ansatz used for their derivation.
Other assumptions to derive \avNcoll, which are discussed in \cite{alice-cent}, use the proportionality 
of \avNcoll to the yield of high-\pt charged particles ($10<p_T<20$~GeV/c) at mid-rapidity
or to the charged-particle multiplicity measured with the V0 detector in the Pb-going direction at forward rapidity.
The variations on the \avNcoll values obtained with the three methods do not exceed 6\% for any of the 
centrality classes used for this analysis and are taken into account as a systematic uncertainty uncorrelated over centrality.
Uncertainties of 8\% and 3.4\% on the determination of \avNcollMB and \avTpaMB, respectively, are also included as global systematic uncertainties. These uncertainties are obtained by varying the parameters of the Glauber model. 
Events without a signal in the ZN detector, which correspond to very peripheral events~\cite{alice-cent}, are assigned to the 
80--100\% centrality interval. 
The values of \avNcollmult and \avTpamult used in 
this analysis\footnote{The 0-2\% ZN class is excluded in the dimuon analysis due to significant pile-up contribution as detailed in Section~\ref{sec:analysismumu}.} are reported in Tab.~\ref{tab:Ncoll}, together with 
their uncertainties.

\begin{table*}[thb!f]
 \centering
\begin{tabular}{c|c|c}
\hline
 ZN class & \avNcollmult  & \avTpamult \\
\hline
 2--10\%   & $11.7 \pm 1.2 \pm 0.9 $  & $0.167 \pm 0.012\pm 0.006  $\\
 10--20\%  & $11.0 \pm 0.4 \pm 0.9$  & $0.157 \pm 0.006 \pm 0.005$\\
 20--40\%  & $9.6 \pm 0.2 \pm 0.8$  & $0.136 \pm 0.003 \pm 0.005$\\
 40--60\%  & $7.1 \pm 0.3 \pm 0.6$  & $0.101 \pm 0.005 \pm 0.003$\\
 60--80\%  & $4.3 \pm 0.3 \pm 0.3$ & $0.061 \pm 0.004 \pm 0.002$\\
 80--100\% & $2.1 \pm 0.1 \pm 0.2$  & $0.030 \pm 0.001 \pm 0.001$\\
\hline
0--20\% & $11.4 \pm 0.6 \pm 0.9$ & $0.164 \pm 0.009 \pm 0.006$\\
60--100\% & $3.2 \pm 0.2 \pm 0.3$ & $0.046 \pm 0.002 \pm 0.002$ \\
\hline
\end{tabular}
 \caption{ Average numbers of binary nucleon-nucleon collisions \avNcollmult
   and average values of the nuclear overlap function \avTpamult with
   their uncorrelated and global systematic uncertainty for the used centrality classes.
   The centrality intervals are expressed as percentages of the non-single diffractive \pPb cross section.
 \label{tab:Ncoll}} 
\end{table*}

\section{Analysis in the dimuon decay channel}
\label{sec:analysismumu}
The analysis approach and the selection criteria are similar to those described in detail in Ref.~\cite{ALICEpA:2013}. 
The primary vertex is reconstructed from the hits in the SPD. No specific requirement is applied on the vertex properties. 
In pile-up events, the ZN energy of the two (or more) interactions are summed up, increasing the pile-up event contribution in the most central ZN class. 
This contribution is estimated to be large, of 
the order of 20--30\%, for events belonging to the centrality class 
0--2\% and this class has therefore been discarded from the analysis. 
The muon candidate tracks are reconstructed in the muon spectrometer tracking stations 
using the algorithm described in Ref.~\cite{Aamodt:2011gj}. 
In order to remove particles at the edge of the muon spectrometer acceptance, a fiducial cut on the single-muon 
pseudorapidity $-4<\eta_{\rm lab}<-2.5$ is applied. 
An additional selection on the radial coordinate of the track at the exit of the front absorber ($17.6<R_{\rm abs}<89.5$ cm) 
is required to reject muons crossing the high-density section of the absorber, where energy loss and multiple 
scattering effects play an important role. Finally, only tracks matching the corresponding tracks reconstructed in the trigger stations are selected. 

\begin{figure}[!tbp]
\begin{center}
\includegraphics[width=1.\linewidth,keepaspectratio]{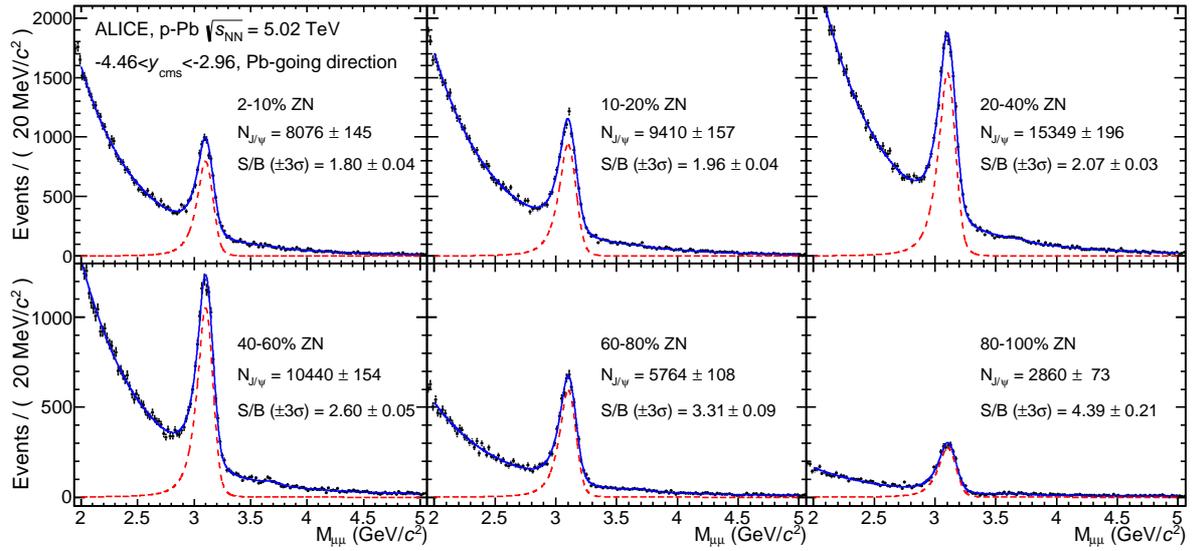}
\caption{(Colour online) Unlike-sign dimuon invariant mass distributions for six \centrality classes 
for $\pt < 15$ \gevc in the \Pbp configuration. The solid blue curves correspond to a fit based on a pseudo-Gaussian function 
for the \jpsi and \psip signals (see text) and an exponential multiplied by a second order polynomial function for the background. 
The red dashed lines represent the fitted signal function.}
\label{fig:SignalExtraction}
\end{center}
\end{figure}

The \jpsi candidates are obtained by combining pairs of muons of opposite charge that are reconstructed in the rapidity range $2.5<|\ylab|<4$ and with $\pt < 15$~\gevc. 
The raw \jpsi yield is estimated for each centrality and \pt interval from fits of the dimuon invariant-mass distribution performed with various functions. 
For the signal component, an extended Crystall Ball function, 
which includes non-Gaussian tails on either side of the \jpsi peak, as well as a pseudo-Gaussian function with a 
mass-dependent width~\cite{Lakomov:publicnote} are employed. 
Due to the poor signal-to-background (S/B) ratio in the tail regions, Monte Carlo (MC) simulations 
are used to constrain the tail parameters in each \pt and rapidity interval under study. 
Since there is no degradation of the tracking resolution due to the large occupancy corresponding to the most 
central \pPb collisions~\cite{Abelev:2014ffa}, the tails are not expected to depend on centrality.
The \psip resonance is also included in the fit function using the strategy described in Ref.~\cite{Abelev:2014qha}. 
For the background component, two alternative functions are used: 
a Gaussian with a mass-dependent width and an exponential multiplied by a second-order polynomial. 
The fits are performed in two different invariant mass ranges, $2<\mmumu<5$~\gevcsqr and 
$2.3<\mmumu<4.7$~\gevcsqr. 
In the fitting procedure, the mean and width of the \jpsi signal function, the background parameters 
and the normalisation factors are left free while the tail parameters are fixed to the values estimated from the simulations. 
 The obtained \jpsi mass value agrees with the 
PDG value~\cite{Agashe:2014kda} within 5~MeV/$c^{2}$. The measured width increases from 59 to 81~MeV/$c^{2}$ with increasing \pt. 
It is found to be about 10\% larger than in the simulations. 
The S/B ratio in the $3\sigma$\ interval around the \jpsi pole increases with increasing \pt 
and towards peripheral events, ranging from 1.1 (1.3) to 5.2 (9) in the \Pbp (\pPb) configuration. 
Figure~\ref{fig:SignalExtraction} shows examples of fits to the unlike-sign dimuon pair invariant mass 
distributions for the \Pbp configuration for six \centrality classes for $\pt < 15$~\gevc. 

The invariant mass fits are performed using the different combinations of signal and background functions and 
fitting ranges described above. The number of \jpsi is obtained by integrating the \jpsi signal function over the fitting range. 
The mean of the distribution of the number of \jpsi obtained from the various fits 
is used as the central value of the raw yield, while the Root Mean Square (RMS) is used as a systematic uncertainty, which ranges between 
0.2\% and 3\% depending on \pt and \centrality class. 
An additional systematic uncertainty of 2\% is added to the signal extraction uncertainty. It is estimated 
from the variation of the raw yield when performing the fit with different tail parameters of the signal function. 
The raw \jpsi yield
varies between about $3000$ and $16000$ counts for \Pbp and between about $5000$ and $17000$ 
counts for \pPb in the 
\centrality-differential results. In the case of the \centrality- and \pt- double-differential results, it 
varies between about $70$ and $4000$ counts in \Pbp 
and between about $150$ and $4000$ counts in \pPb, where 
the lower values correspond to the centrality class 80--100\% and the highest \pt range.  

The \jpsi raw yields are corrected for the detector acceptance and efficiency 
($A\times\epsilon$) estimated from simulations of the \jpsi signal. 
The muon decay products of the \jpsi are propagated through the experimental 
setup modeled with GEANT 3.21~\cite{Brun:1994aa}. 
The procedure used for track reconstruction is the same in data and simulations.
In the latter, the detector conditions and their variation with time during the data-taking period
are taken into account.
It was checked that the detector occupancy in central collisions does not deteriorate 
the single muon tracking efficiency and resolution, 
which justifies that only the \jpsi signal is simulated and not the underlying \pPb collision. 
The \pt and rapidity distributions of the \jpsi signal in the simulation 
were tuned to the reconstructed distributions of the \pPb and \Pbp data using an iterative procedure.
The \jpsi production is assumed to be unpolarised, consistent with the observation that  
no significant \jpsi polarisation has been measured in \pp collisions at 
$\sqrt{s}$~=~7~TeV~\cite{Abelev:2011md, Aaij:2013nlm, Chatrchyan:2013cla}. 
The values of $A\times\epsilon$ integrated over \pt are 17.1\% and 25.4\% in the \Pbp and \pPb configurations, respectively. 
The lower $A\times\epsilon$ for the \Pbp configuration is due to a smaller 
detector efficiency in the corresponding data-taking period. 
The $A\times\epsilon$ varies as a function of \pt from $16\%$ to $33\%$ in \Pbp and from 
$23\%$ to $48\%$ in \pPb collisions, 
where the lowest values correspond to $1<\pt<2$~\gevc and the largest to $8<\pt<15$~\gevc.
The systematic uncertainty in the choice of the \jpsi kinematic distributions in the simulation is estimated 
by varying the \jpsi \pt and rapidity distributions according to the measured ones over various sub-ranges of 
$y$, \pt and centrality (see~\cite{ALICEpA:2013} for more details). 
When integrated over \pt, this uncertainty amounts to $1.5\%$ for both the \Pbp and \pPb configurations, 
while for the \pt-differential studies it does not exceed $1.4\%$. 
The uncertainty on the dimuon tracking efficiency amounts to $6\%$ ($4\%$) for \Pbp (\pPb). 
It is evaluated using the difference between the single-muon 
tracking efficiency obtained from simulations and a data-driven approach based on the 
redundancy of the muon tracking stations~\cite{Abelev:2014ffa}, assuming that the efficiencies 
of the two muons are uncorrelated.
This uncertainty is correlated over centrality and is taken as constant as 
a function of \pt. 
The uncertainty on the determination of the dimuon trigger 
efficiency has three contributions, which are correlated over centrality. 
The first uncertainty is due to the statistical 
uncertainty of the 
trigger detector efficiency which is estimated using data. It 
is independent of \pt and amounts to $2\%$. 
The second uncertainty is extracted from the differences observed
between data and simulations for the measured trigger response function in the region close to the trigger threshold.
This uncertainty varies between $0.5\%$ and $3\%$ and is larger at low \pt. 
The third uncertainty is due to the small fraction of 
opposite-sign pairs which are misidentified as like-sign by the trigger system and increases from 0.5 to 3\% with increasing \pt. 
An additional systematic uncertainty results from the choice of the $\chi^2$ cut, which is applied 
to the matching of tracks reconstructed in the muon tracking and trigger system. 
Applied to the number of dimuon pairs, this uncertainty amounts to $1\%$ and is correlated over centrality.

The normalisation factor of dimuon- to MB-triggered events, \fnorm, which is needed to evaluate the integrated luminosity, 
is determined in a two-step procedure 
as the product $F_{2\mu/1\mu} \cdot F_{1\mu/MB}$, where $F_{2\mu/1\mu}$ ($F_{1\mu/MB}$) 
is the inverse of the probability of 
having dimuon-triggered events (single-muon-triggered events) in a corresponding 
data sample of single-muon-triggered events (MB-triggered events). 
The various quantities are estimated from the number of recorded triggered events in each \centrality class. 
This factor can also be obtained from the centrality-integrated value scaled by $N^{\rm cent}_{\rm MB}/N_{\rm MB}$ and 
$N_{\rm DIMU}/N^{\rm cent}_{\rm DIMU}$, the fraction of MB events and the inverse of the fraction of dimuon events in a given 
centrality class, respectively. 
The latter method, which is statistically more accurate, is used for the evaluation of \fnorm. The 
systematic uncertainty is evaluated from the comparison between the two methods and amounts to $1-2\%$ depending on the centrality class. 
The value of \fnorm depends on 
the centrality class and it smoothly increases from $260$ and $3340$ in \Pbp and from $660$ and $3290$ in \pPb from central to 
peripheral collisions. 
The pile-up event contribution to the number of MB events is estimated by using alternatively the 
interaction vertices reconstructed with the SPD to select pile-up events, or a fast simulation describing the ZN energy distribution. 
The pile-up event contribution is larger in the 2--10\% centrality class where it amounts to $3.5\%$ and $2.7\%$ 
in the \Pbp and \pPb beam configurations, respectively. It decreases to less than 2\% in all other centrality classes. It has been included in the systematic uncertainty of \fnorm. It was further checked by using the fast 
simulation that the overall effect of pile-up events, including the shift of events from a given centrality class to a 
more central one, is covered by the systematic uncertainties quoted for pile-up events.

In order to quantify the nuclear effects in \pPb collisions, reference measurements in \pp collisions at the 
same energy are needed. Since there are no experimental data available on the \jpsi production cross section in 
\pp collisions at \s~=~5.02~TeV,
the procedures described in Ref.~\cite{ALICE:2013spa} for the \pt-integrated case and in Ref.~\cite{Adam:2015iga} for the 
\pt-differential case are used. These procedures involve an interpolation in energy and an extrapolation 
in rapidity and are based on existing measurements in \pp collisions at different energies. 
The resulting values of the \jpsi cross section interpolated to $\s=5.02$~TeV are also reported in those references. 

\section{Analysis in the dielectron decay channel}
\label{sec:analysisee}

The analysis method and the selection criteria are similar to those described in detail in Ref.~\cite{Adam:2015iga}. 
Events are selected which contain a primary vertex determined from tracks reconstructed
in the TPC and the ITS. The vertex position is required to have a distance to the nominal IP smaller than 10~cm along the beamline. 
Due to the lower interaction rate in the data sample used for the mid-rapidity analysis, the pile-up event contribution 
is negligible and the events belonging to the ZN centrality class 0--2\% are included in the analysis. 
The electron and positron candidate tracks are reconstructed in the pseudorapidity range $|\eta_{\rm lab}|<0.9$. 
Tracks are required to have at least 70 out of a maximum of 159 clusters in the TPC, 
a $\chi^2$ normalised to the number of attached TPC clusters smaller than 4 and a distance of closest 
approach to the primary vertex smaller than 3~cm 
along the beam axis and 1~cm in the plane transverse to the beam axis. 
Only tracks with one associated track point in the innermost layer of the ITS and 
at least a second one in the other layers are selected. 
This requirement suppresses the secondary electrons from photon conversions in the detector material of the ITS. 
Tracks are required to be compatible within $3.0\sigma$ with the electron hypothesis based on 
the measured ionisation energy loss 
of the TPC. In order to reject hadrons, the tracks which are compatible within $3.5\sigma$ with the pion 
or proton hypotheses are excluded. 
Tracks from identified photon conversions are rejected without any impact on the \jpsi signal efficiency.
Finally, tracks are required to have a transverse momentum larger than $1$~\gevc in order to improve the S/B 
ratio in the \jpsi mass region~\cite{Aamodt:2011gj}. 

\begin{figure}[!tbp]
\begin{center}
\includegraphics[width=1.\linewidth,keepaspectratio]{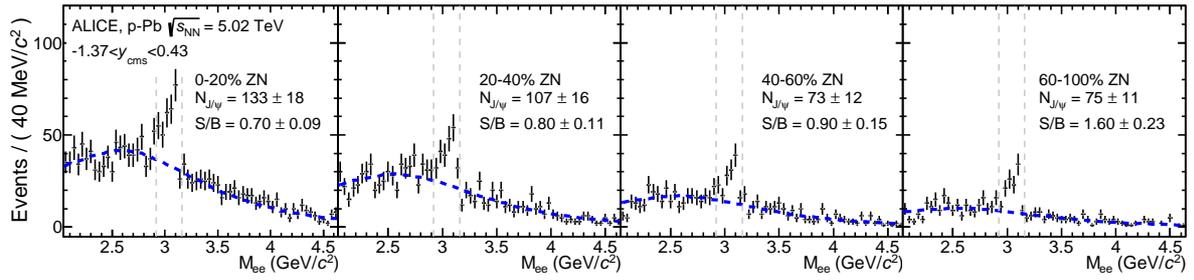}
\caption{(Colour online) Invariant mass distributions of unlike-sign \elel pairs at mid-rapidity for the four considered 
\centrality classes. The background shape, represented by the dashed blue line, is derived from the event mixing technique. The two vertical dashed lines shown 
in each panel indicate the invariant mass interval used for the signal counting.
}
\label{fig:Signalextractionee}
\end{center}
\end{figure}

The \jpsi signal is extracted from the invariant mass distribution of \elel candidates. 
The raw \jpsi yields are estimated 
by bin counting in the invariant mass range $2.92-3.16$~GeV/$c^2$. 
The combinatorial background is estimated using the event mixing technique, 
i.e. by pairing electrons and positrons from different events. 
The event mixing is performed in classes of events, sorted according to multiplicity and vertex position. 
The invariant mass distribution of \elel pairs obtained from this procedure is normalised to the integral of the same-event 
unlike-sign dielectron pairs in the mass ranges $2.0-2.5$~GeV/$c^2$ and $3.2-3.7$~GeV/$c^2$, outside of the signal counting
interval. 
A significant fraction of the \jpsi yield, determined from simulations to be about $30\%$, falls outside
the signal counting window, and is corrected for.  
This is due to the long tail at low masses caused
by the bremsstrahlung of electrons in the detector material and by the radiative soft photon that may 
be emitted at the \jpsi decay vertex. 
The systematic uncertainty on the signal extraction procedure, including uncertainties on the mixed-event background scaling
and on the invariant-mass shape of the \elel decay channel, is obtained by varying the mass region 
used for the scaling of the mixed-event background and by varying the mass window used for counting the signal. 
Figure~\ref{fig:Signalextractionee} illustrates the signal extraction procedure for the \centrality classes considered in this analysis. 
The central value of the \jpsi raw yield is obtained as the average of the raw yields retrieved from the variation of 
the signal extraction configurations described above and 
its systematic uncertainty is the RMS of the distribution of the extracted signals. 
The raw yield has also been evaluated using the like-sign method where the residual background after the like-sign subtraction is estimated by a linear function and the signal is extracted by bin counting. 
The two methods have been found to provide results compatible within the estimated uncertainties. 
As a function of centrality, the \jpsi raw yield varies from 73 to 133 counts, with S/B in the interval 
$2.92 < m_{\elel} < 3.16$ \gevcsqr ranging from 0.7 to 1.6. 

The $A\times\epsilon$ correction is 
estimated with simulations consisting of \jpsi particles 
set to decay into an \elel pair added to a \pPb event generated using HIJING~\cite{Wang:1991hta}. 
The produced particles are subsequently propagated through the experimental setup 
modeled using GEANT 3.21~\cite{Brun:1994aa} and the same reconstruction procedure as for the real data is followed.
 The simulated \jpsi mesons are assumed to be unpolarised. 
The average value obtained for $A\times\epsilon$ is $7.2\%$ with no observed dependence on the collision \centrality
but with a significant dependence on the \jpsi \pt, having a maximum of $\sim11\%$ at zero \pt, a minimum
of $\sim6\%$ at 2~\gevc, a second maximum of $\sim10\%$ at 7~\gevc followed by a slow decrease towards higher momenta.
This shape is due to the kinematic selections and the momentum dependence of the particle identification selection efficiency.

The uncertainty on the \jpsi reconstruction efficiency is dominated by 
the uncertainty on the electron identification. It is estimated 
from the difference in the TPC specific energy loss distribution of a clean sample of electrons from identified 
photon conversions in data and electrons from the simulation.
For dielectron pairs, this uncertainty amounts to $4\%$ and is 
correlated over centrality.
Since the $A\times\epsilon$ is not constant as a function of \pt, its \pt-integrated 
value depends on the \pt shape used to generate the \jpsi mesons in the simulation. To estimate the uncertainty due to the
generated \pt shape, the \pt-differential \jpsi spectrum is varied in the simulations.
The \pt distribution is parameterised as 
\begin{equation}
\label{eq:fpt}
f(\pt)=C \frac{\pt}{(1+(\pt/p_0)^2)^n},
\end{equation}
where $C$, $p_0$ and $n$ are parameters constrained by the experimental results in Ref.~\cite{Adam:2015iga}.
The systematic uncertainty is estimated from the variation of the \pt-integrated $A\times\epsilon$ when varying the fit 
parameters within their uncertainties. It amounts to $3\%$ and is correlated among the \centrality classes. 
Furthermore, an uncertainty due to the variation of the spectral shape 
as a function of \centrality is also taken into account, given the fact that statistical uncertainties 
do not allow for a double-differential measurement. 
To evaluate this effect, the relative variation of the fit parameters of the \pt distributions measured at forward and backward rapidity for several centrality classes is used.
  A maximum variation of the $A\times\epsilon$ by $1.4\%$ is observed and assigned 
as an uncorrelated uncertainty over \centrality.

The inclusive \jpsi production cross section in pp collisions at $\s=5.02$~TeV, which is needed to quantify the nuclear effects in p--Pb collisions, is obtained using the interpolation procedure described in Ref.~\cite{Adam:2015iga}. The method is based on existing measurements in pp collisions at different energies. 

\section{Results}
\label{sec:results}

The double-differential \jpsi production cross section for a given
centrality class is
\begin{equation}
\label{eq:sigma}
\frac{ {\rm d^2} \sigma^{\rm cent}_{\jpsi}}{{\rm d}y{\rm d}\pt}=
\frac{Y^{\rm cent}_{\jpsitoll}}
{ {\rm BR} }
\times \sigmb,
\end{equation}
where \sigmb is the \pPb(\Pbp) MB cross section discussed in Sec.~\ref{sec:genandcent},
 ${\rm BR}$ is the branching 
ratio of the considered \jpsi dileptonic decay channel, which amounts 
to $(5.96 \pm 0.03) \%$ and $(5.97 \pm 0.03) \%$ 
for the dimuon and the dielectron decay channels~\cite{Agashe:2014kda}, respectively, and $Y^{\rm cent}_{\jpsitoll}$ is 
the inclusive \jpsi yield per-event. 
The latter is defined as
\begin{equation}
\label{eq:invy}
Y^{\rm cent}_{\jpsitoll}= 
\frac{N_{\jpsitoll}}{\nmb \cdot (A \times \epsilon) \cdot \Delta y \cdot \Delta \pt},
\end{equation}
where $N_{\jpsitoll}$ is the raw number of \jpsi mesons decaying into dileptons for a given \centrality class, 
rapidity and \pt range, $N_{\rm MB}$ is the number of MB events for 
the given \centrality class, $A \times \epsilon$ is the acceptance times efficiency described in 
Sec.~\ref{sec:analysismumu} and~\ref{sec:analysisee} and $\Delta y$ and $\Delta \pt$ are the widths of the rapidity 
and \pt intervals, respectively. 
Table~\ref{tab:syst} gives a summary of the systematic uncertainties of the \jpsi differential cross section, as well as 
the correlations of these uncertainties over \centrality, collision system and \jpsi \pt.
The \pt-integrated \jpsi cross sections 
are reported in Tab.~\ref{tab:xsec} for the three rapidity intervals 
as a function of \centrality expressed in percentiles of the non-single diffractive 
\pPb cross section. 
\begin{table}[h]
\begin{center}
\small\addtolength{\tabcolsep}{3pt}
\begin{tabular}{c|c|c|c}
\hline
\multirow{2}{*}{Source of uncertainty} & $-4.46<y_{\rm cms}<-2.96$ & $2.03<y_{\rm cms}<3.53$ & $-1.37<y_{\rm cms}<0.43$ \\
                  &  cent. (cent. and \pt) &  cent. (cent. and \pt)  &  cent. \\
\hline
Signal extraction & $2.0-2.4\%$ ($2.8-7.1\%$)  & $2.0-2.1\%$ ($2.1-5.3\%$)  &  $3.7-7.4\%$      \\    
$\mu^{+}\mu^{-}$ tracking (I)      & 6\%                    & 4\%                    & -               \\
$\mu^{+}\mu^{-}$ trigger (I)       & 3.4\% ($2.7-3.6\%$)      & 3\% ($2.7-3.6\%$)      & -               \\
$\mu^{+}\mu^{-}$ matching (I)      & 1\%                    & 1\%                    & -               \\
$e^{+}e^{-}$ reconstruction (I)& -                      & -                      & 4\%             \\
MC input (I)      & 1.5\% ($0.1-1.4\%$)      & 1.5\% ($0.1-0.4\%$)      &   3\%         \\
MC input          & -                      &   -                    &   1.4\%         \\
\fnorm (III)      & $1-3.5\%$                & $1-2.7\%$                &  -               \\
\hline
\multicolumn {4}{c}{Uncertainties related to cross section only} \\
\hline
$\sigma_{\rm MB}$ (I,II,III) & 1.6\%        &  1.6\%                 &  1.6\%          \\
$\sigma_{\rm MB}$ (I,III)    & 3\%          & 3.3\%                  &  3.3\%           \\ 
BR (I, II, III)   & 0.5\%                  & 0.5\%                  & 0.5\%              \\
\hline
\multicolumn {4}{c}{Uncertainties related to \Qpa only} \\
\hline
\avTpPbmult (I,II,III) & 3.4\%             & 3.4\%                  &  3.4\%          \\
\avTpPbmult (II,III)   & $1.9-7.2\%$         & $1.9-7.2\%$              & $1.9-5.6\%$         \\
\sigpp (I)             & 5.3\% ($8.1-13\%$)& 5.7\% ($8.2-11\%$)     &     17\%       \\
\sigpp (I, II, III)    & 5.5\%             & 5.5\%                  &        -         \\
\hline
\end{tabular}
\end{center}
\caption{Summary of the relative systematic uncertainties for the differential \jpsi cross section and \Qpa. 
In the backward and forward rapidity intervals, the uncertainties for the \pt-differential case are indicated in parentheses if different from the \pt-integrated case. 
Type I stands for uncertainties correlated over \centrality. 
Type II corresponds to the uncertainties correlated between the rapidity intervals. 
Type III is related to forward and backward rapidity intervals only and represents the \pt-correlated uncertainties. 
The uncertainty on MC input, $\sigma_{\rm MB}$, \avTpPbmult and \sigpp are split into 
different components according to their correlations over centrality, rapidity intervals and \jpsi \pt.
}
\label{tab:syst}
\end{table}

\begin{table}[h]
\centering
\small\addtolength{\tabcolsep}{3pt}
\begin{tabular}{ccc|cc}
\hline
\multirow{2}{*}{ZN class}  & 
\multicolumn{2}{c|}{${\rm d}\sigma^{\rm cent}_{\jpsi}/{\rm d}y$ ($\mu$b)} 
& \multirow{2}{*}{ZN class} & ${\rm d}\sigma^{\rm cent}_{\jpsi}/{\rm d}y$ ($\mu$b)\\ 
  & $-4.46<\ycms<-2.96$ & $2.03<\ycms<3.53$ & & $-1.37<\ycms<0.43$ \\
\hline
2--10\%  & $1185 \pm 20 \pm 49 \pm 94$ & $944 \pm 16 \pm 33 \pm 61$ & 0--20\% &  $1582\pm 236\pm 120\pm 98 $\\
10--20\% & $1109 \pm 18 \pm 32 \pm 88$ & $885 \pm 14 \pm 25 \pm 57$  & 20--40\% & $1331\pm 204\pm 55\pm 83$\\
20--40\% & $894 \pm 11 \pm 23 \pm 71$  & $811 \pm 10 \pm 21 \pm 53$  & 40--60\% & $890\pm 160\pm 42\pm 55$\\
40--60\% & $617 \pm 9 \pm 15 \pm 49$ &  $603 \pm 8 \pm 15 \pm 39$ & 60--100\% & $460\pm 70\pm 52\pm 29$\\
60--80\% & $330 \pm 6 \pm 10 \pm 26$ & $385 \pm 6 \pm 11 \pm 25$  & & \\
80--100\% & $175 \pm 4 \pm 4 \pm 14$ & $220 \pm 5 \pm 5 \pm 14$   & & \\
\hline
\end{tabular}
\caption{Differential cross sections as a function of \centrality. The first quoted uncertainty is statistical 
while the second and third represent the systematic uncertainties, the latter being fully correlated over \centrality.}
\label{tab:xsec}
\end{table}

\begin{figure}[!tbp]
	\centering
	\includegraphics[width=0.95\columnwidth]{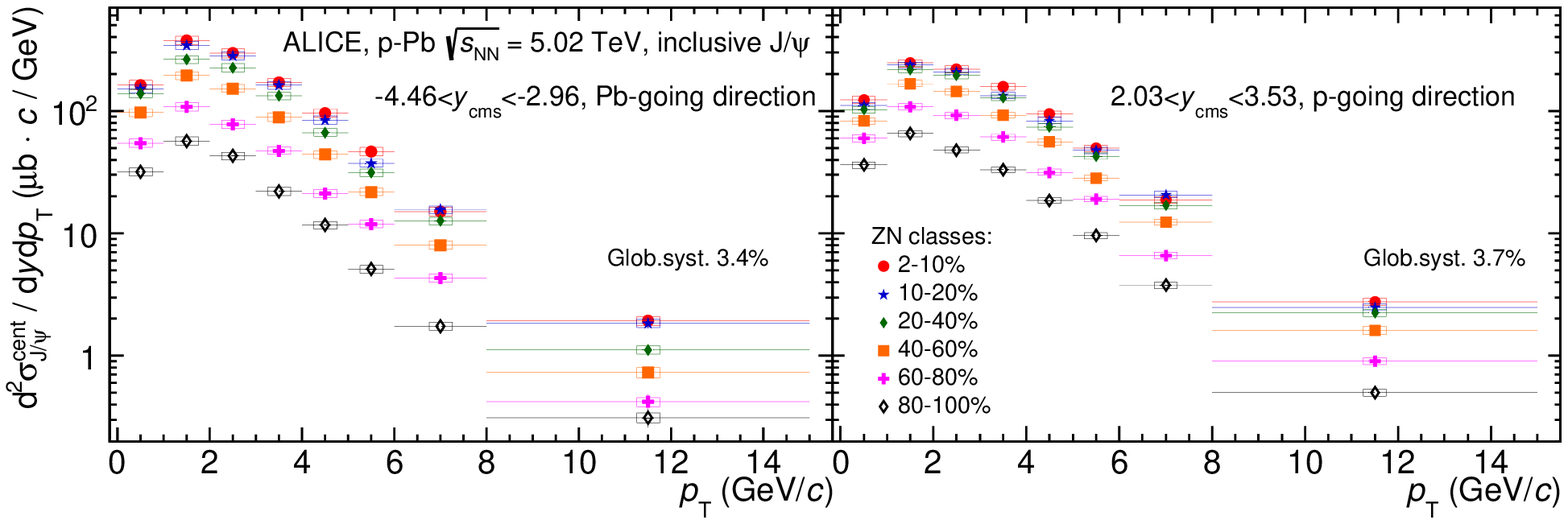}
	\caption{(Colour online) Inclusive \jpsi double-differential cross sections as a function of \pt for various \centrality classes at backward (left) 
and forward (right) rapidity. The systematic uncertainties correlated over \centrality and \pt 
are indicated as a global systematic uncertainty.}
        \label{fig:dsigmadydpt}
\end{figure}

Figure~\ref{fig:dsigmadydpt} shows the double-differential \jpsi cross sections as a function of \pt in the range $0<\pt<15$~\gevc 
at backward (left panel) and forward (right panel) rapidity measured for six \centrality classes. The vertical error 
bars represent the statistical uncertainties and the open boxes the systematic uncertainties. 
The systematic uncertainties correlated over \centrality and \pt are indicated as a global relative systematic uncertainty.
In order to characterise the evolution of the \pt-differential cross section shape with \centrality,
the average values \meanpt and \meanptsqr, 
were extracted for each \centrality class by performing a fit to the data with the function defined in Eq.~\ref{eq:fpt}.
The systematic uncertainties on the data points that are correlated over \pt are 
not considered in the fit. The uncertainties on the free parameters obtained from the fit of Eq.~\ref{eq:fpt} 
are propagated to the values of \meanpt and \meanptsqr. The statistical and systematic uncertainties on 
\meanpt and \meanptsqr are obtained by performing the fit using, separately, 
only the statistical or the uncorrelated systematic uncertainties on the data points, respectively. 
The range of integration over \pt used to compute \meanpt and \meanptsqr 
is limited to the measured \pt interval $0<\pt<15$~\gevc. It was verified that 
extending the integration range to infinity results in an increase of \meanpt and \meanptsqr 
values by less than 0.5\%. 
The values of \meanpt and \meanptsqr obtained for each \centrality class are reported 
in Tab.~\ref{tab:meanpt}. 
Both \meanpt and \meanptsqr values increase with \centrality, 
which indicates a hardening of the \pt distributions from peripheral to central collisions in both rapidity intervals. 
\begin{table}[h!]
\begin{center}
\small\addtolength{\tabcolsep}{3pt}
\begin{tabular}{c|cc|cc}
\hline
\multirow{2}{*}{ZN class} & \multicolumn{2}{c|}{$-4.46<y_{\rm cms}<-2.96$}  & \multicolumn{2}{c}{ $2.03<y_{\rm cms}<3.53$} \\
 & \meanpt (\gevc) & \meanptsqr (\gevcsqr) & \meanpt (\gevc) & \meanptsqr (\gevcsqr) \\
\hline
2--10\% & 2.53 $\pm$ 0.02 $\pm$ 0.03 & 9.12 $\pm$ 0.17 $\pm$ 0.23   & 2.82 $\pm$ 0.03 $\pm$ 0.03 & 11.43 $\pm$ 0.21 $\pm$ 0.23 \\
10--20\% & 2.52 $\pm$ 0.02 $\pm$ 0.03 & 9.14 $\pm$ 0.16 $\pm$ 0.23 & 2.85 $\pm$ 0.03 $\pm$ 0.03 & 11.66 $\pm$ 0.20 $\pm$ 0.24 \\
20--40\% & 2.48 $\pm$ 0.02 $\pm$ 0.04 & 8.74 $\pm$ 0.12 $\pm$ 0.22 & 2.81 $\pm$ 0.02 $\pm$ 0.03 & 11.36 $\pm$ 0.15 $\pm$ 0.23 \\
40--60\% &  2.44 $\pm$ 0.02 $\pm$ 0.04 & 8.48 $\pm$ 0.14 $\pm$ 0.22 & 2.75 $\pm$ 0.02 $\pm$ 0.03 & 10.98 $\pm$ 0.16 $\pm$ 0.24 \\
60--80\% & 2.41 $\pm$ 0.03 $\pm$ 0.04 & 8.41 $\pm$ 0.20 $\pm$ 0.22 & 2.65 $\pm$ 0.02 $\pm$ 0.03 & 10.22 $\pm$ 0.18 $\pm$ 0.21 \\
80--100\% & 2.32 $\pm$ 0.03 $\pm$ 0.04 & 8.03 $\pm$ 0.28 $\pm$ 0.22 & 2.61 $\pm$ 0.03 $\pm$ 0.03 & 10.00 $\pm$ 0.24 $\pm$ 0.22 \\
\hline
\pp     & $2.37\pm0.04$ & $8.18\pm0.30$ & $2.52\pm0.04$ & $9.28\pm0.40$ \\
\hline
\end{tabular}
\caption{Values of \meanpt and \meanptsqr of inclusive \jpsi in the range $0<\pt < 15$ \gevc. The first quoted uncertainty is statistical while the 
second is systematic. The values obtained from the \pp cross section interpolated to \s~=~5.02~TeV
are also indicated.}
\label{tab:meanpt}
\end{center}
\end{table}

\begin{figure}[!tbp]
	\centering
	\includegraphics[width=0.8\columnwidth]{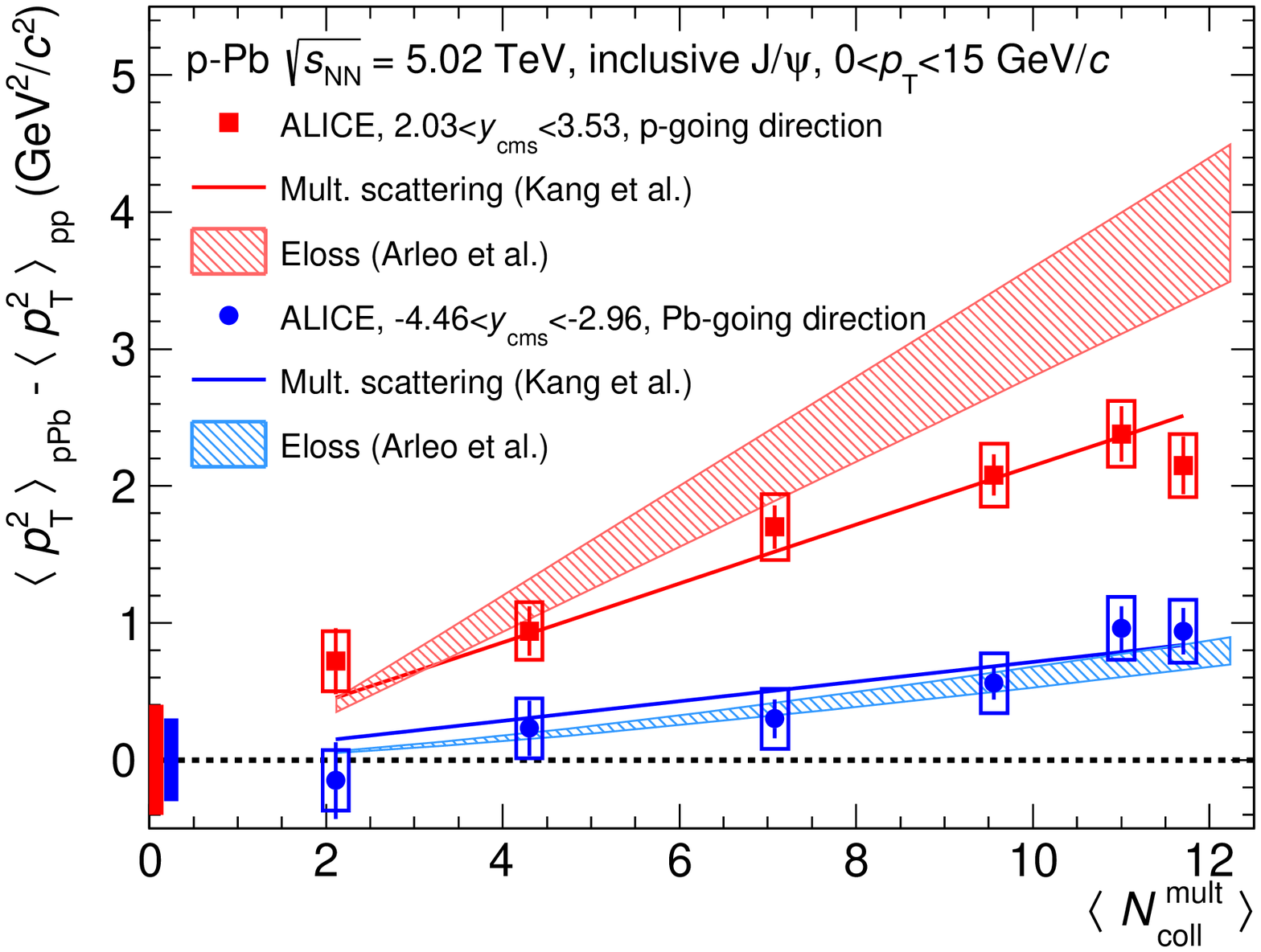}
	\caption{(Colour online) \pt broadening, \deltameanptsqr, as 
a function of \avncollmult at backward (blue circles) and forward (red squares) rapidity. 
The boxes centered at $\deltameanptsqr=0$ represent the total uncertainties of $\meanptsqr_{pp}$ 
interpolated to $\s=5.02$~TeV. 
The theoretical calculations are from Refs.~\cite{Kang:2012am,Kang:2008us,Arleo:2013pt}.}
\label{fig:deltameanptsqr}
\end{figure}

In order to quantify the nuclear effects on the \jpsi \pt spectrum shape, the 
\pt broadening, \deltameanptsqr, defined as
\begin{equation}
\label{eq:deltapt}
\deltameanptsqr = \meanptsqr_{\mathrm{pPb}} - \meanptsqr_{\mathrm{pp}},
\end{equation}
is used. 
Since there are no measurements for \pp collisions at \s~=~5.02~TeV, 
the value of $\meanptsqr_{\mathrm{pp}}$ is evaluated from the \pt-differential cross section in \pp collisions calculated 
with the interpolation procedure described in Ref.~\cite{Adam:2015iga}, and using the same \pt-integration range as for \pPb collisions. 
Figure~\ref{fig:deltameanptsqr} shows \deltameanptsqr as a function of the number of binary collisions. 
Our measurements indicate that \deltameanptsqr increases at backward (forward) rapidity 
by $\sim1.1$~$\mathrm{GeV^2}/c^{2}$ ($\sim1.4$~$\mathrm{GeV^2}/c^{2}$) from peripheral to central \pPb collisions. At forward rapidity, 
\deltameanptsqr is larger for all centrality classes and suggests a steeper dependence on centrality compared 
to backward rapidity values.  
For the most peripheral collisions, corresponding to $\ncollmult\sim2$, 
the \meanptsqr value at backward rapidity is compatible with the one in \pp collisions, while at forward rapidity it is found to be 
larger than in pp collisions by $0.7$~$\mathrm{GeV^2}/c^2$, which corresponds to 1.4 times the total
uncertainty on the measured difference. 
The 
magnitude of the \pt broadening observed by PHENIX~\cite{Adare:2012qf} in \dAu collisions at $\snn=200$~GeV in the rapidity ranges 
$-2.2<\ycms<-1.2$, $|\ycms|<0.35$ and $1.2<\ycms<2.2$ is similar to the one measured by ALICE at backward rapidity. 
At forward rapidity, the ALICE data show a stronger \pt broadening and a steeper increase with increasing centrality 
as compared to PHENIX results. 

The calculations from Refs.~\cite{Kang:2012am,Kang:2008us} are based on the leading order (LO) CEM  
production model and include initial and final-state multiple 
scattering of partons with the nuclear medium (denoted as \quotes{Mult. scattering} in Fig.~\ref{fig:deltameanptsqr}). 
The uncertainties on the theoretical calculations are not available. In this model, the 
contribution to \pt broadening due to final-state multiple scattering 
is expected to be sensitive to the colour-octet or colour-singlet nature of the pre-resonant \ccbar pair.  
The calculations are in good agreement with the data at backward and forward rapidity. 
A second model, which is based on a parameterisation of the prompt \jpsi \pp cross section
and includes coherent energy loss effects from the incoming and outgoing partons~\cite{Arleo:2013pt} 
(denoted as \quotes{Eloss} in Fig.~\ref{fig:deltameanptsqr}), 
describes well the centrality dependence of \deltameanptsqr at backward rapidity. 
The trend predicted by this model at forward rapidity is slightly steeper than the data even considering 
its uncertainty, evaluated by varying the gluon transport coefficient 
and the parametrisation of the production cross section.  

In order to study the modification of the \jpsi production 
in \pPb collisions with respect to \pp interactions, 
the \jpsi nuclear modification factor is used. For a given \centrality class, rapidity and \pt range, it is defined as
\begin{equation}
\label{eq:qpa}
Q^{\mathrm{mult}}_\mathrm{pPb} = \frac{Y^{\mathrm{cent}}_{\jpsitoll}}
{\avTpPbmult \cdot BR \cdot {\rm d}^2\sigpp/{\rm d}y{\rm d}p_{\rm T} }.
\end{equation}
The \avTpPbmult values corresponding to the \centrality 
classes used in this analysis are reported in Tab.~\ref{tab:Ncoll}. 
The \jpsi cross section in \pp collisions at $\s=5.02$~TeV, ${\rm d}^2\sigpp/{\rm d}y{\rm d}p_{\rm T}$, 
is obtained by means of the interpolation procedures outlined in Sec.~\ref{sec:analysismumu} and~\ref{sec:analysisee} and described in Refs.~\cite{ALICE:2013spa,Adam:2015iga}.
The nuclear modification factor is usually denoted as \Rpa but in this analysis the notation \Qpa is used to emphasise 
the possible bias in the evaluation of \avTpPbmult, as discussed in Sec.~\ref{sec:genandcent} and in Ref.~\cite{alice-cent}. 
The systematic uncertainties on \Qpa are presented in Tab.~\ref{tab:syst}. 

\begin{figure}[t!]
\begin{center}
	\includegraphics[width=1.\linewidth,keepaspectratio]{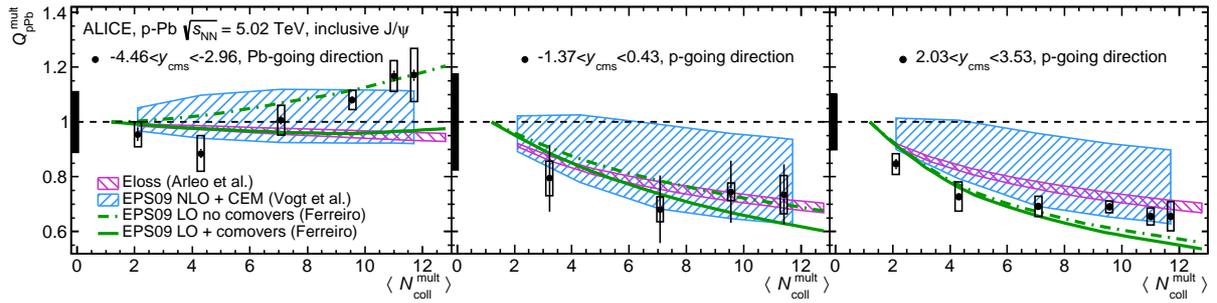}
	\caption{(Colour online) Inclusive \jpsi \Qpa as a function of \avncollmult at backward (left), mid (center) and 
forward (right) rapidity. The boxes centered at \Qpa = 1 represent the relative uncertainties correlated over \centrality. The theoretical calculations are from Refs.~\cite{Vogt:2010aa,Vogt:2013,Ferreiro:2014bia,Arleo:2013pt}.}
	\label{fig:qpa}
\end{center}
\end{figure}

Figure~\ref{fig:qpa} shows the dependence of 
the \pt-integrated \Qpa on the collision centrality, expressed as \avncollmult. The results for the backward, mid- and forward rapidity 
   intervals are displayed in the left, middle and right panel, respectively.
 At backward rapidity, the \Qpa values show that the measured \jpsi production is compatible, within the total uncertainties, 
with expectations from binary collision scaling for all \centrality classes. When considering only the uncertainties that are not correlated over centrality, an increase from peripheral to central \pPb collisions is observed in the data. 
At forward rapidity, the \jpsi yield is suppressed with respect to the binary-scaled \pp reference 
for all the considered centrality classes. 
The values of \Qpa measured at forward rapidity exhibit a decrease from $0.85$ for the 80--100\% centrality class down to $0.66$ for the 2--10\% centrality class.
Within the present uncertainties, the mid-rapidity results suggest a similar degree of suppression of the \jpsi yield as at forward rapidity and no conclusion can be drawn on a possible centrality dependence. 

Our measurements are compared to several theoretical models including 
a next-to-leading order (NLO) 
CEM calculation~\cite{Vogt:2010aa,Vogt:2013} which contains the EPS09 NLO nPDF parameterisation~\cite{Eskola:2009uj} 
(denoted as \linebreak[4] \quotes{CEM+EPS09 NLO} in Fig.~\ref{fig:qpa}), a model employing the EPS09 LO nPDF with or without effects from the interaction with 
a comoving medium~\cite{Ferreiro:2014bia} (denoted as \quotes{EPS09 LO+comovers} in Fig.~\ref{fig:qpa}), and the coherent energy loss model~\cite{Arleo:2013pt} 
described above. 
In the CEM+EPS09 NLO and EPS09 LO+comovers models, 
assuming the \jpsi production process is $gg\rightarrow\jpsi$ ($2\rightarrow1$), 
the \xbj values of the gluon from the Pb nucleus span a range of about 
$1\cdot10^{-2}<\xbj<5\cdot10^{-2}$ at backward rapidity, $4\cdot10^{-4}<\xbj<2\cdot10^{-3}$ 
at mid-rapidity, 
and $2\cdot10^{-5}<\xbj<8\cdot10^{-5}$ at forward rapidity. The backward rapidity interval therefore corresponds 
to the \xbj range in the transition between the anti-shadowing and the shadowing region, 
whereas the mid- and forward rapidity intervals 
probe a region for which the gluon shadowing is expected to be strong. 
The CEM+EPS09 NLO model uncertainties are evaluated from the EPS09 uncertainty, which gives the dominant contribution,
 and from a variation of the values of the charm quark mass, the normalisation and the factorisation scales in the pQCD calculation. 
The CEM+EPS09 NLO model reproduces well the centrality 
dependence in each rapidity range. At mid- and forward rapidity, the data are better reproduced when a strong shadowing is considered in the model. 
In the framework of the EPS09 LO+comovers model, the presence of a comoving medium has only a small 
effect on \jpsi production at forward rapidity since its density is expected to decrease towards the p-going direction. 
The effect of comovers is more pronounced 
at mid-rapidity and especially at backward rapidity and it increases with increasing centrality. 
The uncertainties on these theoretical calculations are not available. 
At backward rapidity, the 
increase of \Qpa towards central collisions 
observed in the data  
is better reproduced when the comover effect is not included in the model. 
Finally, the shape and magnitude of \Qpa is well described by the Eloss model in all rapidity intervals, although 
the model does not predict an increase with increasing centrality at backward rapidity, 
as indicated by the data. 

\begin{figure}[!t]
\begin{center}
	\includegraphics[width=0.8\linewidth,keepaspectratio]{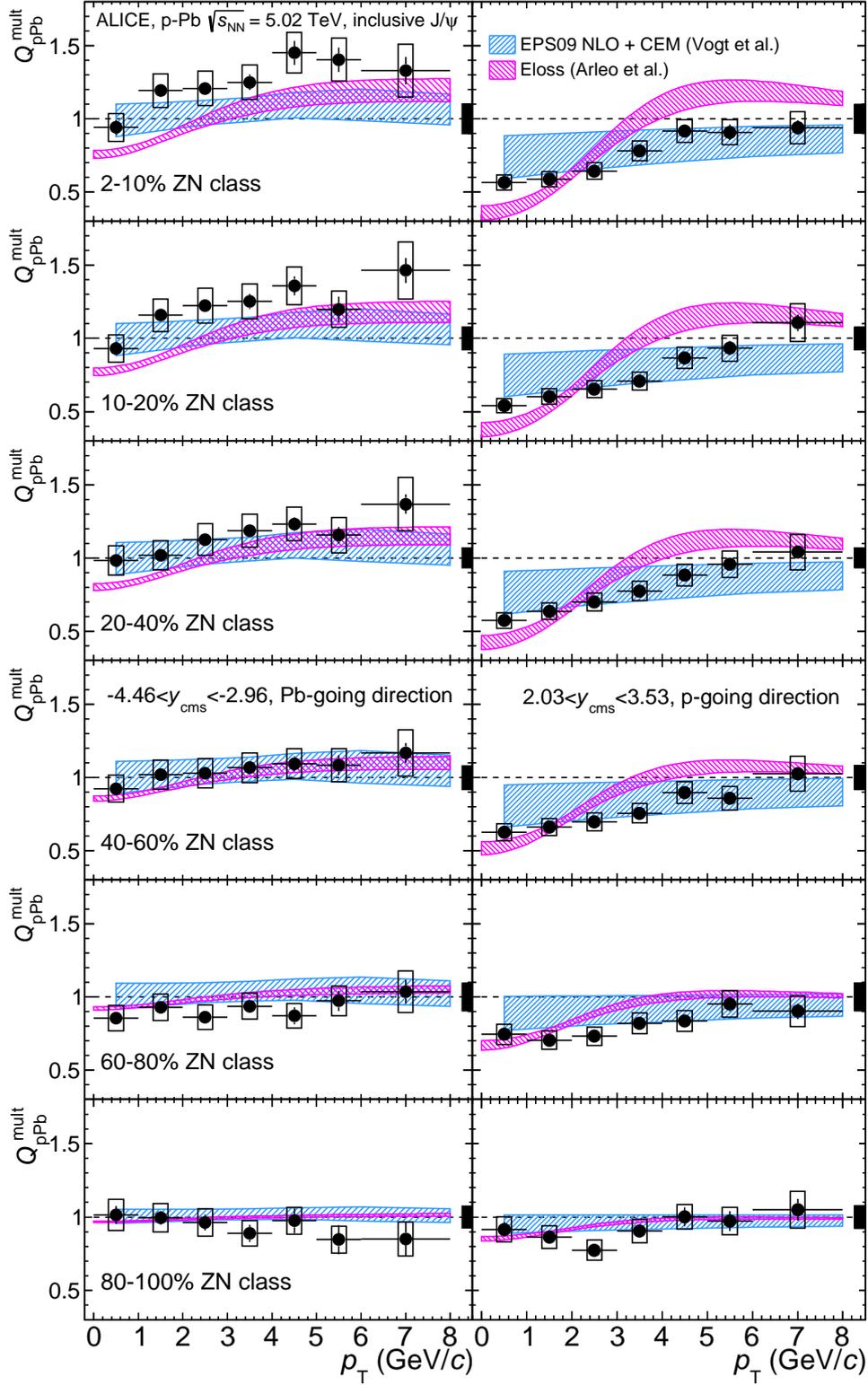}
	\caption{Inclusive \jpsi \Qpa as a function of \pt for the 2--10\%, 10--20\%, 20--40\%, 40--60\%, 60--80\% and 80--100\% (from top to bottom) ZN \centrality classes at backward (left) and forward (right) rapidity. The boxes centered at \Qpa~=~1 represent the relative uncertainties correlated over \pt. The theoretical calculations are from Refs.~\cite{Vogt:2010aa,Vogt:2013,Arleo:2013pt}.}
        \label{fig:dqpadpt}
\end{center}
\end{figure}
It is worth pointing out that the calculations above are done for prompt \jpsi production, while the 
measurements also include the contribution of \jpsi mesons from $b$-hadron decays. 
The \Qpap can be extracted 
from \Qpainc using the relation $\Qpap = \Qpainc + \fb \cdot (\Qpainc - \Qpab)$, 
where \fb is the ratio of non-prompt to prompt \jpsi production cross sections 
and \Qpab is the nuclear modification 
factor of non-prompt \jpsi mesons. 
A value of \fb of about 0.11 at $2<\ycms<4.5$  
and for $\pt<14$~\gevc can be calculated 
from the LHCb measurements in \pp collisions at \s~=~7~TeV~\cite{LHCbpp7:2011}. 
The value of \fb does not show a strong variation  
within the quoted rapidity range and with energy, as indicated 
by the comparison with the results in \pp collisions at \s~=~8~TeV~\cite{LHCbpp8:2013}. Hence, the value of \fb 
calculated at \s~=~7~TeV in $2<\ycms<4.5$ is used for the following. 
At mid-rapidity, a value of \fb of about 0.17 at $|y|<0.9$ and integrated over \pt can be extracted 
from the measurements of ALICE in \pp collisions at \s~=~7~TeV~\cite{Abelev:2012gx}. 
The nuclear modification factor of non-prompt \jpsi was measured 
to be $0.98\pm0.06\pm0.01$ ($0.83\pm0.02\pm0.08$) for $-4<\ycms<-2.5$
($2.5<\ycms<4$) and $\pt < 14$~\gevc at \snn~=~5.02~TeV in \pPb collisions~\cite{Aaij:2013zxa}. 
If the non-prompt \jpsi \Qpa, which has not been measured as a function of the centrality, 
is conservatively assumed to vary from 0.6 to 1.3 in each centrality interval, then the differences between the inclusive 
and prompt \jpsi nuclear modification factors cannot exceed 15\% in any of the centrality classes and are smaller than 
the quoted uncertainties. 

Figure~\ref{fig:dqpadpt} presents the \pt-dependence of \Qpa for the 2--10\%, 10--20\%, 20--40\%, 40--60\%, 
60--80\% and 80--100\% \centrality classes, from the top to the bottom panels, respectively. The left (right) panels show 
the backward (forward) rapidity results. 
At backward rapidity and for the most central collisions (2--10\% and 10--20\% \centrality classes), 
\Qpa is compatible with unity in the full \pt interval, and an increase of 
\Qpa from $\pt<1$~$\mathrm{GeV}/c$ to $\pt>1$~$\mathrm{GeV}/c$ is suggested by the data. 
For semi-central and peripheral collisions, 
\Qpa is compatible with unity over the full \pt range. 
At forward rapidity, 
for all \centrality classes other than the most peripheral one, \jpsi production is 
suppressed compared to binary-scaled \pp production at low \pt. For these centrality classes, \Qpa increases with increasing 
\pt and is compatible with unity at high \pt within the uncertainties. The magnitude of \Qpa 
slightly increases from central to semi-peripheral collisions over the full \pt range. 
For the most peripheral collisions \Qpa is compatible with unity and, with the current uncertainties,
it does not show a significant \pt dependence. 

The data are compared to the calculations from the CEM+EPS09 NLO~\cite{Vogt:2010aa,Vogt:2013} and the 
Eloss~\cite{Arleo:2013pt} model. 
The CEM+EPS09 NLO calculations describe reasonably well the \Qpa results 
at backward and forward rapidity. 
The Eloss model reproduces well the \pt dependence of \Qpa at backward rapidity for all centrality 
classes. 
At forward rapidity, a good agreement of the calculations with the data is observed for peripheral collisions 
(60--80\% and 80--100\% centrality classes), while the \pt dependence becomes steeper than in data towards more central collisions. 
It was observed in Ref.~\cite{Adam:2015iga} 
that a better agreement is reached with the \pt dependence of the nuclear modification factor in centrality-integrated \pPb collisions
at forward rapidity when shadowing effects are included in the model. 

\section{Conclusions}

The cross sections and nuclear modification factors, \Qpa, of inclusive \jpsi production 
have been measured with ALICE as a function of rapidity, \pt and centrality
in \pPb collisions at \snn~=~5.02~TeV. 
For the most peripheral \pPb collisions, no modification with respect to \pp collisions 
is observed within the uncertainties of the measurements for both the shape of the \pt spectrum of the \jpsi  
and the \Qpa measurements. 
On the contrary, the results in central \pPb collisions suggest sizeable  
nuclear effects. At both backward (Pb-going direction) and forward (p-going direction) 
rapidity the \deltameanptsqr measurements show a \pt broadening 
which increases monotonically from peripheral to central \pPb collisions with larger values 
at forward rapidity. 
Our measurements show a stronger \pt broadening and a steeper increase with increasing centrality at forward rapidity 
as compared to PHENIX results in \dAu collisions at $\snn=200$~GeV~\cite{Adare:2012qf}. 
At backward rapidity, a modest increase of \jpsi production compared to a binary-scaled \pp reference is suggested 
by the data in most central collisions. 
At mid-rapidity, the data indicate that \jpsi production is suppressed compared to binary-scaled \pp cross section 
over the entire \centrality range. Within the current uncertainties, the increasing suppression towards central \pPb 
collisions suggested by models is compatible with the data. 
Finally, at forward rapidity, a clear suppression, which increases towards central events, is observed. 
The \pt- and \centrality-differential results show that the suppression is
stronger at low \pt and tends to vanish at high \pt. 
Given the uncertainties of both the measurements and the theoretical calculations, we observe a fair agreement of the models based on coherent energy 
loss and multiple scattering with the measured \pt broadening. Models based on nPDF and coherent energy loss are in fair agreement 
with the nuclear modification factor measurements. 
The results presented in this paper provide an important baseline for understanding and constraining the cold 
nuclear matter effects in \pPb collisions as well as their 
centrality dependence. 
Such an information is essential for a quantitative interpretation of the results obtained in \PbPb collisions. 

\newenvironment{acknowledgement}{\relax}{\relax}
\begin{acknowledgement}
\section*{Acknowledgements}
The ALICE Collaboration would like to thank all its engineers and technicians for their invaluable contributions to the construction of the experiment and the CERN accelerator teams for the outstanding performance of the LHC complex.
The ALICE Collaboration gratefully acknowledges the resources and support provided by all Grid centres and the Worldwide LHC Computing Grid (WLCG) collaboration.
The ALICE Collaboration acknowledges the following funding agencies for their support in building and
running the ALICE detector:
State Committee of Science,  World Federation of Scientists (WFS)
and Swiss Fonds Kidagan, Armenia,
Conselho Nacional de Desenvolvimento Cient\'{\i}fico e Tecnol\'{o}gico (CNPq), Financiadora de Estudos e Projetos (FINEP),
Funda\c{c}\~{a}o de Amparo \`{a} Pesquisa do Estado de S\~{a}o Paulo (FAPESP);
National Natural Science Foundation of China (NSFC), the Chinese Ministry of Education (CMOE)
and the Ministry of Science and Technology of China (MSTC);
Ministry of Education and Youth of the Czech Republic;
Danish Natural Science Research Council, the Carlsberg Foundation and the Danish National Research Foundation;
The European Research Council under the European Community's Seventh Framework Programme;
Helsinki Institute of Physics and the Academy of Finland;
French CNRS-IN2P3, the `Region Pays de Loire', `Region Alsace', `Region Auvergne' and CEA, France;
German Bundesministerium fur Bildung, Wissenschaft, Forschung und Technologie (BMBF) and the Helmholtz Association;
General Secretariat for Research and Technology, Ministry of
Development, Greece;
Hungarian Orszagos Tudomanyos Kutatasi Alappgrammok (OTKA) and National Office for Research and Technology (NKTH);
Department of Atomic Energy and Department of Science and Technology of the Government of India;
Istituto Nazionale di Fisica Nucleare (INFN) and Centro Fermi -
Museo Storico della Fisica e Centro Studi e Ricerche "Enrico
Fermi", Italy;
MEXT Grant-in-Aid for Specially Promoted Research, Ja\-pan;
Joint Institute for Nuclear Research, Dubna;
National Research Foundation of Korea (NRF);
Consejo Nacional de Cienca y Tecnologia (CONACYT), Direccion General de Asuntos del Personal Academico(DGAPA), M\'{e}xico, :Amerique Latine Formation academique – European Commission(ALFA-EC) and the EPLANET Program
(European Particle Physics Latin American Network)
Stichting voor Fundamenteel Onderzoek der Materie (FOM) and the Nederlandse Organisatie voor Wetenschappelijk Onderzoek (NWO), Netherlands;
Research Council of Norway (NFR);
National Science Centre, Poland;
Ministry of National Education/Institute for Atomic Physics and Consiliul Naţional al Cercetării Ştiinţifice - Executive Agency for Higher Education Research Development and Innovation Funding (CNCS-UEFISCDI) - Romania;
Ministry of Education and Science of Russian Federation, Russian
Academy of Sciences, Russian Federal Agency of Atomic Energy,
Russian Federal Agency for Science and Innovations and The Russian
Foundation for Basic Research;
Ministry of Education of Slovakia;
Department of Science and Technology, South Africa;
Centro de Investigaciones Energeticas, Medioambientales y Tecnologicas (CIEMAT), E-Infrastructure shared between Europe and Latin America (EELA), Ministerio de Econom\'{i}a y Competitividad (MINECO) of Spain, Xunta de Galicia (Conseller\'{\i}a de Educaci\'{o}n),
Centro de Aplicaciones Tecnológicas y Desarrollo Nuclear (CEA\-DEN), Cubaenerg\'{\i}a, Cuba, and IAEA (International Atomic Energy Agency);
Swedish Research Council (VR) and Knut $\&$ Alice Wallenberg
Foundation (KAW);
Ukraine Ministry of Education and Science;
United Kingdom Science and Technology Facilities Council (STFC);
The United States Department of Energy, the United States National
Science Foundation, the State of Texas, and the State of Ohio;
Ministry of Science, Education and Sports of Croatia and  Unity through Knowledge Fund, Croatia.
Council of Scientific and Industrial Research (CSIR), New Delhi, India
\end{acknowledgement}

\makeatletter
\bibliographystyle{utphys}
\renewcommand{\@biblabel}[1]{#1.}
\makeatother

\bibliography{./biblio.bib}

\addcontentsline{toc}{chapter}{\tocsecindent{References}}

\newpage
\appendix

\section{The ALICE Collaboration}
\label{app:collab}
\begin{flushleft} 

\bigskip 

J.~Adam$^{\rm 40}$, 
D.~Adamov\'{a}$^{\rm 83}$, 
M.M.~Aggarwal$^{\rm 87}$, 
G.~Aglieri Rinella$^{\rm 36}$, 
M.~Agnello$^{\rm 111}$, 
N.~Agrawal$^{\rm 48}$, 
Z.~Ahammed$^{\rm 132}$, 
S.U.~Ahn$^{\rm 68}$, 
I.~Aimo$^{\rm 94}$$^{\rm ,111}$, 
S.~Aiola$^{\rm 136}$
% ...

\bigskip

\bigskip 

\textbf{\Large Affiliation Notes}

\bigskip 

% ...
$^{\rm III}$ Also at: University of Kansas, Lawrence, Kansas, United States

\bigskip

\bigskip 

\textbf{\Large Collaboration Institutes}

\bigskip 

% ...
$^{36}$ European Organization for Nuclear Research (CERN), Geneva, Switzerland\\
% ...
$^{40}$ Faculty of Nuclear Sciences and Physical Engineering, Czech Technical University in Prague, Prague, Czech Republic\\
% ...
$^{48}$ Indian Institute of Technology Bombay (IIT), Mumbai, India\\
% ...
$^{68}$ Korea Institute of Science and Technology Information, Daejeon, South Korea\\
% ...
$^{83}$ Nuclear Physics Institute, Academy of Sciences of the Czech Republic, \v{R}e\v{z} u Prahy, Czech Republic\\
% ...
$^{87}$ Physics Department, Panjab University, Chandigarh, India\\
% ...
$^{94}$ Politecnico di Torino, Turin, Italy\\
% ...
$^{111}$ Sezione INFN, Turin, Italy\\
% ...
$^{132}$ Variable Energy Cyclotron Centre, Kolkata, India\\
% ...
$^{136}$ Yale University, New Haven, Connecticut, United States\\
% ...

\bigskip 

\end{flushleft} 

\end{document}